\documentclass[fontsize=6pt,journal=apchd5,manuscript=article]{achemso}
\usepackage[version=3]{mhchem} 
\usepackage[numbers,sort&compress,super]{natbib}
\usepackage{hyperref}
\usepackage[dvipsnames]{xcolor}
\usepackage{bm}
\usepackage{setspace}

\newcommand{\beqa}{\begin{eqnarray}}
\newcommand{\eeqa}{\end{eqnarray}}

\newcommand{\bra}[1]{\langle #1|}
\newcommand{\ket}[1]{|#1\rangle}

\author{Simone Felicetti$^1$}
\affiliation{Istituto di Fotonica e Nanotecnologie, Consiglio Nazionale delle Ricerche (IFN-CNR),\\
Piazza Leonardo da Vinci 32, 20133 Milano, Italy}
\alsoaffiliation{Departamento de F{\'i}sica Te{\'o}rica de la Materia Condensada and Condensed Matter Physics Center (IFIMAC), Universidad Autonoma de Madrid, E-28049 Madrid, Spain}
\author{Jacopo Fregoni$^1$}
\affiliation{Dipartimento di Scienze Chimiche, University of Padova, I-35131 Padova, Italy}
\altaffiliation{Dipartimento di Scienze Fisiche, Informatiche e Matematiche, University of Modena and Reggio Emilia,I-41125 Modena, Italy}
\email{jacopo.fregoni@unimore.it}
\author{Thomas Schnappinger}
\author{Sebastian Reiter}
\author{Regina de Vivie-Riedle}
\affiliation{Department Chemie, Ludwig-Maximilians-Universit\"at M\"unchen,\\ D-81377 M\"unchen, Germany}
\author{Johannes Feist}
\affiliation{Departamento de F{\'i}sica Te{\'o}rica de la Materia Condensada and Condensed Matter Physics Center (IFIMAC), Universidad Autonoma de Madrid, E-28049 Madrid, Spain}
\email{johannes.feist@uam.es}
\title[An \textsf{achemso} demo]{Photoprotecting uracil by coupling with \\
  lossy nanocavities}

\abbreviations{IR,NMR,UV}

\begin{document}
\footnotetext[1]{These authors contributed equally to the realization of the present work}

\newpage

\begin{abstract}
\singlespacing
\noindent We analyze how the photorelaxation dynamics of a molecule can be controlled by modifying its electromagnetic environment using a nanocavity mode. In particular, we consider the photorelaxation of the RNA nucleobase uracil, which is the natural mechanism to prevent photodamage. In our theoretical work, we identify the operative conditions in which strong coupling with the cavity mode can open an efficient photoprotective channel, resulting in a relaxation dynamics twice as fast than the natural one. We rely on a state-of-the-art chemically-detailed molecular model and a non-Hermitian Hamiltonian propagation approach to perform full-quantum simulations of the system dissipative dynamics. By focusing on the photon decay, our analysis 
unveils the active role played by cavity-induced dissipative processes in modifying chemical reaction rates, in the context of molecular polaritonics. Remarkably, we find that the photorelaxation efficiency is maximized when an optimal trade-off between light-matter coupling strength and photon decay rate is satisfied. This result is in contrast with the common intuition that increasing the quality factor of nanocavities and plasmonic devices improves their performance. Finally, we use a detailed model of a metal nanoparticle to show that the speedup of the uracil relaxation could be observed via coupling with a nanosphere pseudomode, without requiring the implementation of complex nanophotonic structures.
\end{abstract}

\begin{figure*}[ht!]
\centering

  \includegraphics[width=10.9cm]{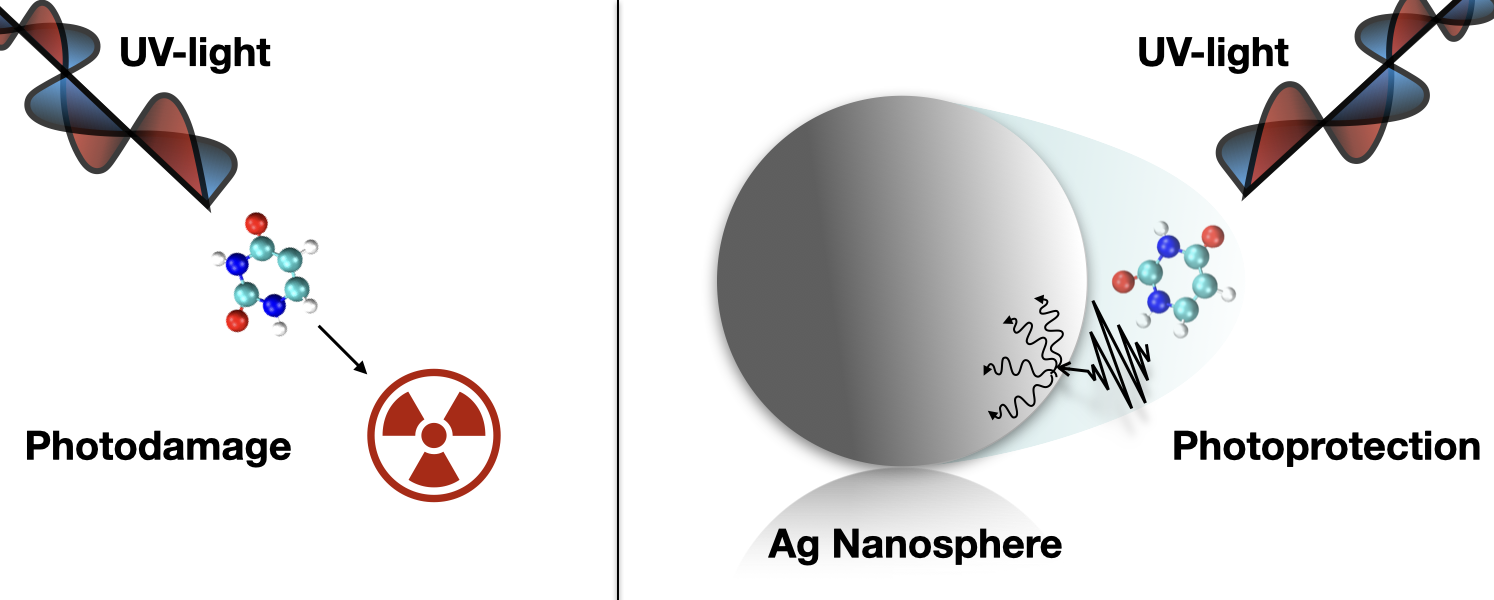}
\end{figure*}


\onehalfspacing
\subsection*{Introduction}

The functionalisation of nanoparticles with macromolecular structures like DNA and RNA is quickly gathering interest as a tool for sensing applications and drug-delivery in medicine and biology\cite{dnaapp,fluorescencedna,rnasensing,rnananop}. When functionalised, the nanoparticles can travel within the biological system and be transported to a target region. Upon irradiation, the functionalized nanoparticles boost the optical response of the surrounding molecules. However, such enhancement does not come without risk: as a result of the electromagnetic field enhancement effect induced by the nanoparticle, the nucleobases in DNA and RNA are more prone to absorb radiation. In turn, the enhanced light absorption can potentially drive the nucleobases to trigger photodamaging chain events\cite{phototoxicdye,goldnpdamage,dnadamage}, resulting in dangerous mutations of the DNA/RNA macrostructure. Although characterized by an intrinsically-low quantum yield, the pathway associated to photodamage occurs as a side intersystem crossing reaction starting from the $S_2$ bright excited states of nucleobases\cite{dnaphotoform,doorway,dnadamage}. Under ambient conditions of irradiation, this low quantum yield is mostly due to the quick de-excitation mechanism of the nucleobases. This mechanism then prevents dangerous reactions\cite{barbatti:bases}, with an estimated permanence on the $S_2$ excited state of $\sim$0.5 ps for both DNA\cite{barbatti:bases,doorway,dna:repair} and RNA\cite{leticia:uracil,regina1,regina2}. However, this self-preservation mechanism fails under under high-intensity irradiation, like in fluorescence imaging techniques or \textit{in-vitro} preparations\cite{livecell:photo,biobook,rnadamage,rnashadowing}. It then becomes natural to look for new pathways to improve the photoprotection mechanism, already at the single-molecule level.

A compelling strategy to control the chemical and dynamical properties of molecules relies on the modulation of their electromagnetic environment via cavity quantum electrodynamics (cQED) devices.
The collection of models and experimental techniques aiming at controlling chemistry via coupling to quantum light
goes by the name of polaritonic chemistry or molecular
polaritonics\cite{feist:polchem,ebbesen:rev,rubio:review,ribeiro:polchem,borje:rev,herrera:molpol}. 
In this framework the molecules are confined in optical cavities and are resonantly coupled to localized modes of the electromagnetic field \cite{mukamel:cavity,spano:coherent,galego:cavity,ebbesen:dynamics}.
Whether this coupling is strong enough to drive substantial
chemical modifications depends on the
specifics of the system\cite{flick:weaktostrong}: the oscillator strength of the 
molecular excitation, the volume of the mode in the
nanocavity, the number of molecular emitters, together with
the lifetimes of the exciton and nanocavity mode. When the
coupling strength exceeds the decay rates of both the cavity
mode and the exciton (strong-coupling regime), the states of
the system can be suitably described as hybrids between light and matter: the polaritons\cite{herrera:molpol}. 
Very recent theoretical works and experiments have proven that
reaching the strong coupling regime is a useful way to catalyze photochemical reactions\cite{mukamel:novel,ebbesen:switch,feist:manymols,fregoni:chem,ribeiro:singletsplit}, to modify the relaxation pathways of
molecules\cite{dunkel:modifieddyn,vendrell:moddyn,groenh:relax} and to mediate energy transport phenomena\cite{ebbesen:et,pupillo:et,feist:et}, as well
as to enhance the molecular optical response\cite{baumb:lh,coles:chlorosomes,groenh:multiscale2,martinez:harvesting}. In the same fashion, Shegai and collaborators have shown that strong coupling to plasmonic nanoantennas can signifcantly increase the photostability of chromophores\cite{shegai:photostability}.

Promising experimental setups exploited in
molecular polaritonics rely on nanocavities coupled to
organic molecules\cite{specialissue}. This kind of setup is beneficial as it
guarantees high oscillator strengths on the molecular side and
a nanometric mode volume on the nanocavity side\cite{vanhulst:nanocav,baranov:novel,baumberg:dnaorigami}, carrying the
possibility to observe molecular strong coupling down to the
single molecule level at room temperature\cite{baumberg:singlemol,baumberg:qed}. However, it comes
with the drawback of the modes living  for only a few tens of
femtoseconds\cite{baumberg:nanocav2,aizpurua:nanocav}. It is then a common aim in the field to work
towards extending the cavity lifetimes, guided by the
intuitive idea that stronger coupling would correspond to more
exotic properties. While this may be desirable in
light-harvesting\cite{baumb:lh,coles:chlorosomes,groenh:multiscale2,martinez:harvesting} and energy transport applications\cite{ebbesen:et,pupillo:et,feist:et}, we show here that increasing the mode lifetime is not necessarily
the best approach to pursue for opening up efficient
relaxation pathways.

In the present work, we show how the uracil photorelaxation mechanism can be improved by coupling with a lossy nanocavity mode. 
As a starting point, we analyze the photorelaxation dynamics of the isolated uracil molecule. This first step lets us identify how to act with the nanocavity mode on the photoprotection mechanism. 
Utilizing a non-Hermitian formalism to include cavity losses, we perform quantum dynamics simulations and look for the cavity parameters that optimize the photorelaxation mechanism of the isolated nucleobase.
Indeed, we identify such conditions
between coupling strength and mode lifetime and we characterize the mechanism leading to the improved photoprotection.
Our simulations also reveal that the best efficiency is obtained when the ratio between coupling strength and mode lifetime is at the crossover between weak and strong coupling.
We then show that these coupling conditions can be met by coupling the uracil with a spherical silver nanoparticle, surrounded by a dielectric with properties reflecting those of a nanoparticle functionalised with DNA\cite{baumberg:dnaorigami}. We also show that, for the case of a silver nanosphere coupled to uracil, no field enhancement in the nucleobase excitation window is present. Thus, the photoprotection is not compromised by an improved absorption of dangerous radiation. The results presented in the current work are doubly beneficial: on one front, we show how to substantially speed-up the photorelaxation of DNA-like structures with simple nanospheres, without relying on complex nanophotonics setups. In a more general perspective, we demonstrate that better cavities do not necessarily correspond to improved photochemistry. A similar non-Hermitian
scheme to investigate cavity losses in Polaritonic Chemistry has been independently developed by Foley and collaborators\cite{antoniou2020role}. There, they thoroughly investigate the role of cavity losses on the non-adiabatic coupling terms for the azobenzene molecule.

\subsection*{Uracil free Evolution}
\begin{figure*}[t]
\centering

  \includegraphics[width=1\textwidth]{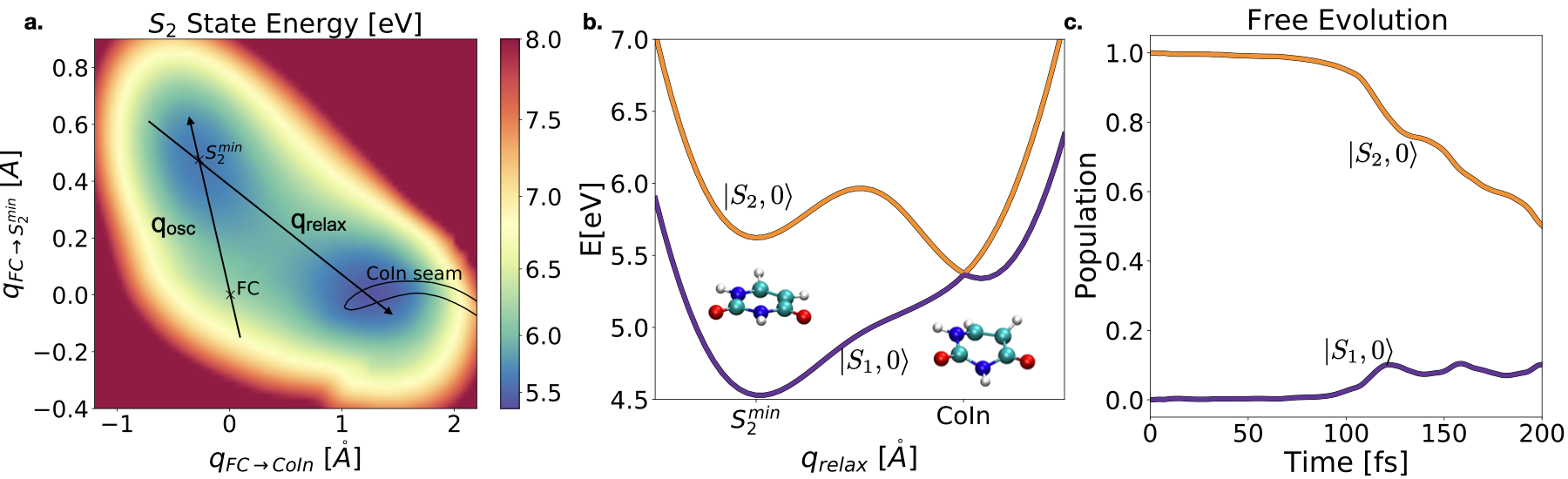}
  \caption{\textbf{Uracil photorelaxation dynamics under free evolution}| \textbf{a)} Main coordinates dominating the photorelaxation in uracil from the $S_2$ state. Absent cavity, the WP oscillates for more than 100 fs along the $q_{osc}$ coordinate, until it manages to overcome the barrier (panel \textbf{b}) along $q_{relax}$. \textbf{c)} Once the WP starts overcoming the barrier, it can reach the $S_2/S_1$ conical intersection and it is quickly transferred to the $\ket{S_1,0}$ state (where $S_1$ denotes the excitonic state, while the second index gives the photon number).}
  \label{fig:coords}
\end{figure*}
Let us first briefly review the relaxation dynamics of the isolated uracil molecule\cite{regina1,regina2}. Uracil, as the other nucleobases, is characterized by a dark $n-\pi^*$ transition ($S_0\rightarrow S_1$) in the visible and a bright $\pi-\pi^*$ ($S_0\rightarrow S_2$) transition occurring by absorption of UV light. The dynamics taking place in the $S_2$ state determines whether the molecule relaxes through internal conversion (photoprotection mechanism) or whether it incurs  photodamage\cite{barbatti:bases,regina1}.
The degrees of freedom responsible for the internal conversion from the $S_2$ excited state involve a collective deformation of the ring-like structure. Although it is known \cite{barbatti:bases,barbatti:nadiab2,mai:bookdna} that an $S_2/S_1$ conical intersection (CoIn) rules the photoprotection mechanism, two more configurations are particularly relevant to properly describe\cite{regina1} the processes occurring in the $S_2$ state: the Franck-Condon point (FC) and a local minimum of $S_2$ ($S_2^{min}$). The $S_2$ potential energy surface (PES) along the vectors ($q_{FC\rightarrow CoIn}$, $q_{FC\rightarrow S_2^{min}}$) respectively connecting the FC to CoIn and to $S_2^{min}$ is displayed in Figure \ref{fig:coords}a. The potential energy landscape presents a double-well structure, with a potential energy barrier hindering the pathway between $S_2^{min}$ and the CoIn seam. We label such pathway as $q_{relax}$ and plot it in Figure \ref{fig:coords}b). 

We compute the PESs on a finite-element discrete-variable (FEDVR)\cite{fedvr} spatial grid as described in the Methods section to represent the nuclear wavefunction.
Upon UV photoexcitation of uracil, the nuclear wavepacket (WP) is transferred from the ground state to the Frank-Condon point (FC) on the $S_2$ state.
Following the shape of the PESs, it then evolves towards $S_2^{min}$ and  starts to oscillate between FC and $S_2^{min}$ along the coordinate labelled as $q_{osc}$. As enough kinetic energy is redistributed during the oscillation to the other coordinate, the WP overcomes the barrier along $q_{relax}$ and it reaches the CoIn seam, finally relaxing to the $S_1$ state. The height of the barrier along $q_{relax}$ is then what determines the permanence time of the wavepacket on the $S_2$ state. Consequently, it is directly related to the probability of triggering a photodamaging reaction (longer permanence time, higher probability to trigger photodamage). A previous work\cite{regina2} also shows how the RNA environment embedding uracil further hinders the relaxation by stabilizing the $S_2^{min}$, resulting in an a higher barrier to overcome and a substantially longer permanence time in S2. In the most optimistic prevision, namely for the isolated molecule, the time needed by the WP to start overcoming the barrier is about 120 fs, as shown in the population plot (Figure \ref{fig:coords}c).

\subsection*{Cavity-assisted photorelaxation}

As the early stages of the relaxation dynamics occur between FC and $S_2^{min}$, an effective way to speed up the dynamics would be to open an alternative relaxation pathway along the $q_{osc}$. The new relaxation channel would be independent both from the height of the barrier and from the environment affecting the depth of $S_2^{min}$. In the course of the present work, we show that such an additional relaxation pathway can be introduced by coupling the molecule with a localized photonic mode. In particular, we consider a single-mode nanocavity whose frequency $\Omega_c$ is set in the near-UV to be resonant with the $S_0\rightarrow S_2$ molecular transition. The cavity-molecule coupling takes advantage of the high transition dipole moment between $S_0$ and $S_2$, resulting in the opening of a direct channel to the ground state. The efficiency of this new channel is strongly affected by photon decay, as discussed later on in the manuscript. In this regard, let us briefly describe how these photon losses can be included in our model.

Quantum systems weakly coupled to dissipative Markovian environments are commonly treated by a Lindblad master equation approach. There, the density matrix of the quantum system is propagated in time, according to eq. \ref{ME} of the Methods section.
In the single-photon subspace, an equivalent method consists in adopting a non-Hermitian formalism, where dissipative effects are implicitly included by adding a complex energy contribution to the lossy states in the effective non-Hermitian Hamiltonian (see Methods section). The photon decay is then accounted for by a loss of norm of the WP during the propagation. For the present case, we include the cavity losses with a non-Hermitian term proportional to the cavity decay rate $\gamma$.

Resorting to the non-Hermitian framework carries two main advantages: first, the photon decay enters the photorelaxation mechanism also at the PESs level via the complex contribution to the energy of lossy states, providing an intuitive picture.
Second, it reduces the computational complexity. On the one hand, it allows us to propagate states instead of the full density matrix. On the other hand, it lets us restrict the state space to the three relevant PESs directly involved in the cavity-assisted photorelaxation. More details are provided in the Methods sections.

\paragraph{Non-Hermitian Polaritonic PESs}
Before moving to the description of the dynamics, we introduce the polaritonic formalism and the relevant states involved in the cavity-assisted photorelaxation mechanism.
\begin{figure*}[t]
\centering
  \includegraphics[width=1\textwidth]{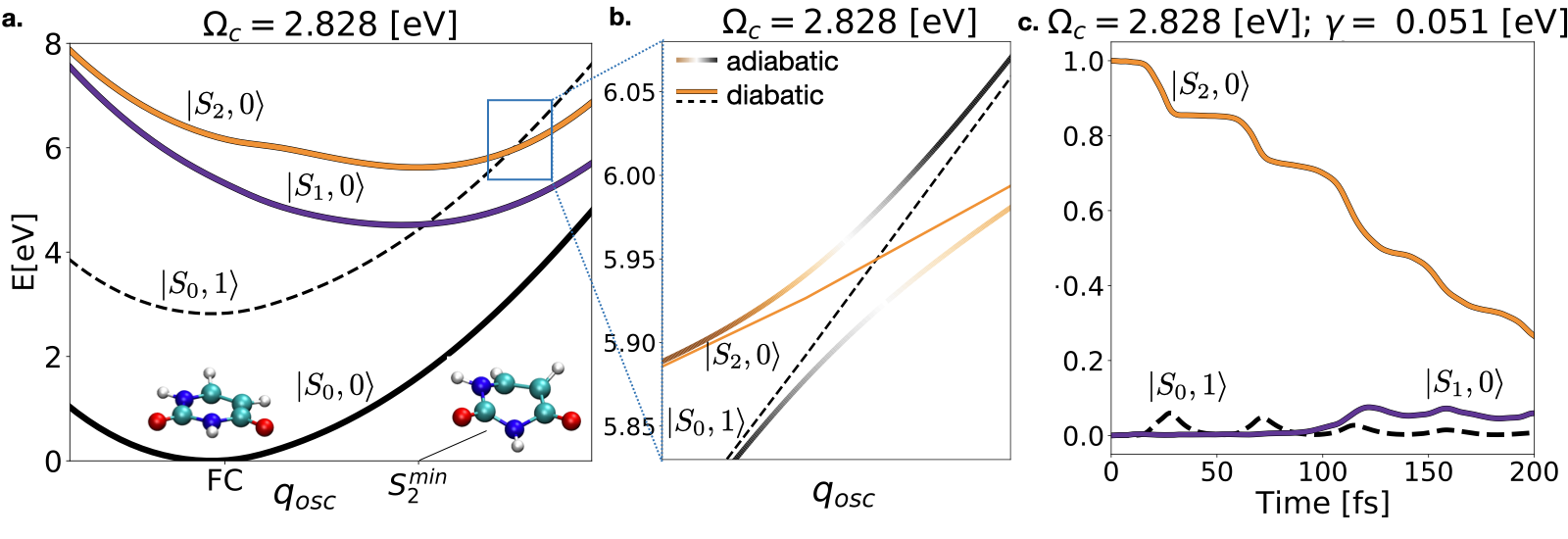}
  \caption{\textbf{Cavity-assisted uracil photorelaxation}|\textbf{a)} Uracil potential energy curves along the $q_{osc}$ coordinate. The full lines represent the isolated molecules potential energy curves, namely $\ket{S_2,0)}$ (orange) $\ket{S_1,0}$ (purple) and $\ket{S_0,0}$ (black). The dashed black line represents the ground state shifted by the cavity mode frequency ($\ket{S_0,1}$). \textbf{b)} Detail of the polaritonic avoided crossing. The diabatic states $\ket{S_2,0}$ and $\ket{S_0,1}$ are split in polaritonic adiabatic states, represented by the shaded lines. \textbf{c)} Population dynamics of uracil excited states in the presence of a cavity mode set at $\Omega_{cav}$ 2.8 eV, $e_{1ph}$ 0.001 au and $\tau_{cav}$ 12 fs. Here, the earlier stage of the dynamics is governed by the new relaxation pathway opened by the presence of the cavity.}
  \label{fig:sc_dynam}
\end{figure*}
In the presence of the cavity, the states associated to the PESs correspond to the tensor product between the electronic states ($\ket{S_i}$) and the cavity mode number states ($\ket{p}$). We label the resulting states as $\ket{S_i,p}$, where $i$ is the electronic state index and $p$ is the cavity-mode occupation number, safely assumed to be either zero or one as the cavity is not externally driven. The energy of the zero-photon states $\ket{S_i,0}$ is purely electronic.
Conversely, when the photon mode is excited, the energy of the states $\ket{S_i,1}$ is 
the $S_i$ PES lifted by the single-photon energy $\Omega_c$. 
The PESs involved in the dynamics are associated to the states $\ket{S_2,0}$, $\ket{S_1,0}$ and $\ket{S_0,1}$.
In the Methods section, we give the detailed derivation of the model and of the space reduction.

To visualize the effects of the cavity on the energy landscape, in Figure \ref{fig:sc_dynam}a we examine the section of the PESs involved in the relaxation along $q_{osc}$. The full lines represent the bare electronic states of the isolated molecule, namely $\ket{S_0,0}$ (black), $\ket{S_1,0}$ (purple) and $\ket{S_2,0}$ (orange). The dashed black line corresponds to the molecule in the ground state and a single photonic excitation, that is $\ket{S_0,1}$. 
By the effect of the strong light-molecule coupling, the bare states $\ket{S_2,0}$ and $\ket{S_0,1}$ hybridize into polaritons, depicted as full shaded lines in Figure \ref{fig:sc_dynam}b. There, the strength $g$ of the coupling term (defined in eq. \ref{lm_coupling} in the Methods section) is given by the product between the molecular transition moment $\mu(\mathbf{R})$ and the single-photon electric field $E_{1ph}$, which depends on the nanocavity design. The polaritonic states are obtained by diagonalisation of the non-Hermitian Hamiltonian, where the energy splitting at the avoided crossings is given by $\sqrt{4 g^2 - \gamma^2}$. The splitting is opened for $g \geq \gamma/2$; this condition is often used to define the onset of the strong coupling regime.

Following the initial oscillation along $q_{osc}$ on the upper polaritonic surface, the composition of the state gradually changes from purely excitonic ($\ket{S_2,0}$) to purely photonic ($\ket{S_0,1}$).
The region of the polaritonic PESs characterized by a major $\ket{S_0,1}$ component (displayed with the colour black) is exposed to photon losses. As a consequence, a WP travelling the black region can incur  an ultrafast radiative de-excitation from $\ket{S_0,1}$ to the $\ket{S_0,0}$ state. The velocity of such de-excitation processes is ruled by the cavity decay rate $\gamma$.
We note for later discussion that the propagation of a nuclear wavepacket following the polaritonic states (shaded lines) is adiabatic. Conversely, we define the motion as diabatic if the wavepacket propagates following the $\ket{S_2,0}$ and $\ket{S_0,1}$ PESs.

\paragraph{Improved photorelaxation of uracil}

\begin{figure*}[t]
\centering
  \includegraphics[width=1.05\textwidth]{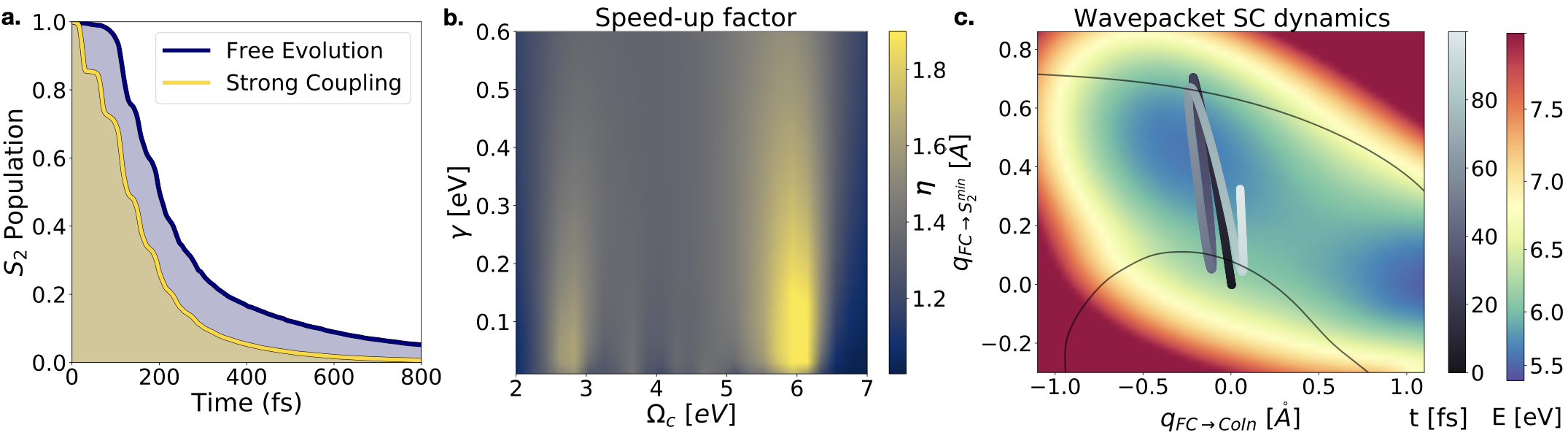}
    \caption{\textbf{Cavity-enhanced photorelaxation mechanism of uracil}| \textbf{a)} Integrated $S_2$ population during the uracil photorelaxation in the presence (yellow) and absence (blue) of a cavity. The total area estimates the probability to incur photodamage. \textbf{b)} Best cavity conditions to optimize the speed-up $\eta$ of the uracil photorelaxation, resulting in a more efficient photoprotection. $\textbf{c)}$ Motion of the WP center of mass. The background shows the $S_2$ PES, while the black lines identifie the seams of intersection between the states $\ket{S_2,0}$ and $\ket{S_0,1}$, for the optimal values $\Omega_c = 2.8$ eV and $\Omega_c = 6.2$ eV of the cavity frequencies. Transfer is maximized if the two states are resonant in the regions where the WP is slower. In general, the conditions leading to the most efficient photoprotection are given by a delicate interplay between cavity frequency, wavepacket velocity and cavity lifetime.}  
  \label{fig:maps}
\end{figure*}

Let us move to the description of the dynamics in the presence of the cavity.
To this aim, we display the populations  for the cavity-assisted dynamics in Figure \ref{fig:sc_dynam}c. Immediately we note that the early dynamics is ruled by the transfer of the WP from $\ket{S_2,0}$ (full orange line) to $\ket{S_0,1}$ (black dashed line), which is active along the $q_{osc}$ coordinate. Here, the state $\ket{S_0,1}$ is only transiently populated, as the WP is quickly lost by the effect of photon leakage, which transfers the WP to the $\ket{S_0,0}$ state. The comparison of the cavity-assisted case to the first 120 fs of the free evolution dynamics (Figure \ref{fig:coords}c) reveals that the cavity  opens a faster and alternative relaxation pathway. Indeed, in the free-evolution case, the WP is still confined at 120 fs by the barrier along $q_{relax}$. The norm of the WP transferred from the $\ket{S_2,0}$ to the $\ket{S_0,1}$ strongly depends on how adiabatic the WP moves on the polaritonic surface, namely on how much it follows the upper polaritonic state of Figure \ref{fig:sc_dynam}b. Indeed, it has recently been noted\cite{rui:clock} that the WP is efficiently transferred with an optimal trade-off between the time it spends in the
coupled region (WP velocity), the speed of the photon leakage (decay rate $\gamma$ or cavity lifetime $\tau_{cav}=1/\gamma$) and how much the states shall be coupled to guarantee an adiabatic motion of the WP from $\ket{S_2,0}$ to $\ket{S_0,1}$.

Aiming to quantify the effective improvement of this new relaxation channel with respect to the different conditions, we adopt as a figure of merit the relaxation speed-up $\eta$. We compute $\eta$ as the ratio between the $\ket{S_2,0}$ population integrated over time for two cases, respectively in the presence of the cavity and for the free evolution (Figure \ref{fig:maps}a). The reason is twofold: firstly, it is directly related to the probability to incur  photodamage. Secondly, it provides a good insight of the overall relaxation process in presence of multiple decay channels acting on different timescales, namely the photon loss and the relaxation through the CoIn seam for the present case. The coupling conditions adopted for the long-time dynamics shown in Figure \ref{fig:maps}a are the same as in Figure \ref{fig:sc_dynam}c, \textit{i.e.} a cavity of $\Omega_c = 2.8$~eV with an associated single-photon field strength of $e_{1ph} = 0.001$~au and a decay rate of about $\gamma$=0.05 eV (lifetime of 12 fs). Under such conditions, the $\ket{S_2,0}$ state is depleted $\sim$1.6 times quicker than for the free evolution case. We now discuss the effect of the coupling conditions on the relaxation dynamics. To this aim, we set the single-photon energy at 0.001 au and consider the speed-up factor for different cavity frequencies and lifetimes. The results are reported in Figure \ref{fig:maps}b. Overall, we obtain a global speed-up factor $\eta>1.2$ for all the range of cavity frequencies between $\Omega_{cav}=$ 2.3 eV and 6.5 eV. 
We identify two regions characterised by a major speed-up factor around two values of the cavity frequency,  $\Omega_{cav}\sim$2.8 eV and $\Omega_{cav}\sim$6 eV.

The cavity frequency rules the position of the polaritonic avoided crossing, and hence the conditions to transfer the WP from $\ket{S_2,0}$ and $\ket{S_0,1}$. By tuning $\Omega_{cav}$, the position of the polaritonic avoided crossing is adjusted along the $q_{osc}$ coordinate. The coupling strength $g$ at the polaritonic avoided crossing (Figure \ref{fig:sc_dynam}b) is consequently affected due to the  dependence of the $S_0\rightarrow S_2$ transition dipole on the nuclear coordinates. However, for the present case, the $S_0\rightarrow S_2$ transition dipole moment is approximatively constant in all the region corresponding to the $S_2$ left minimum\cite{regina1}. Hence, the $\ket{S_0,1}$ state ends up efficiently coupling to $\ket{S_2,0}$ for any $\Omega_{cav}$ in the range between 2.3 eV and 6.5 eV. While this effect explains the overall speed-up of the reaction, it does not account for the maximum speed-up regions at the edges of the frequency range. To understand the optimal speed-up at 2.8 eV and 6.2 eV, we consider the WP oscillation along $q_{osc}$ in Figure \ref{fig:maps}c. Here, we represent the polaritonic avoided crossing in the two cases as black lines labelled by their respective cavity frequencies. We observe that the avoided crossings are located at the edges of the oscillation coordinate $q_{osc}$, where the WP velocity approaches to zero as it reverts its motion. Coherently with the study by Silva \textit{et al.}\cite{rui:clock}, we then see that a more efficient transfer to $\ket{S_0,1}$ is obtained when the WP moves slowly and spends more time in the coupled region, \textit{i.e.} its motion tends to follow the adiabatic behaviour introduced above. The pseudo-harmonic oscillation around the $S_2^{min}$ is also responsible for the stair-like population dynamics of the $\ket{S_2,0}$ state, presented in Figure \ref{fig:sc_dynam}c. In particular, the transfer occurs only when the WP hits the resonant region sitting at the limit of $q_{osc}$.

Considering the WP velocity as a factor affecting the transfer also allows us to understand the dependence of the speed-up factor on the cavity lifetime. Too small decay rate $\gamma$ with respect to the WP permanence time in the coupled region would result in a coherent exchange of the WP back and forth between $\ket{S_2,0}$ and $\ket{S_0,1}$. The coherent exchange continues until the WP exits the coupled region, resulting in an overall minor transfer. On the other hand, a cavity decay rate which is too large with respect to the coupling strength $g$, would result in an effective decoupling of the polaritonic PESs.  
While it is indeed possible to find an optimal $\gamma$ at each cavity frequency, we stress that the effect of $\gamma$ on the speed-up is very dependent on two factors: the coupling strength $g$ and the velocity of the WP when it crosses the polaritonic avoided crossing region. This point will be further discussed for the specific case presented in the following section.

\subsection*{Nanoparticle}
So far, the photonic part of our system has been described with a single-mode cavity coupled to a Markovian dissipation bath. Let us now introduce a more detailed physical model of a nanophotonic structure that reproduces the considered model to a very good degree of approximation. We consider a silver nanoparticle embedded in a nondispersive dielectric medium, and we show that even such a simple structure makes it possible to obtain a significant photorelaxation speed-up with realistic physical parameters. The nanoparticle dielectric response is fitted from the experimental data\cite{jc:diel}, and it is used in a semi-analytic approach\cite{delga:dipole} to evaluate its spectral properties. In particular, we consider a silver nanosphere of radius $a=15$nm, embedded in a high-refractive-index continuous dielectric. We take a value of the background dielectric constant of $\epsilon_d= 4.41$, which has been experimentally observed by Baumberg and collaborators\cite{baumberg:dnaorigami} using DNA-origami. 
We place the molecule at a distance of $1$ nm from the nanoparticle surface.

\begin{figure*}[t]
\centering

  \includegraphics[width=1\textwidth]{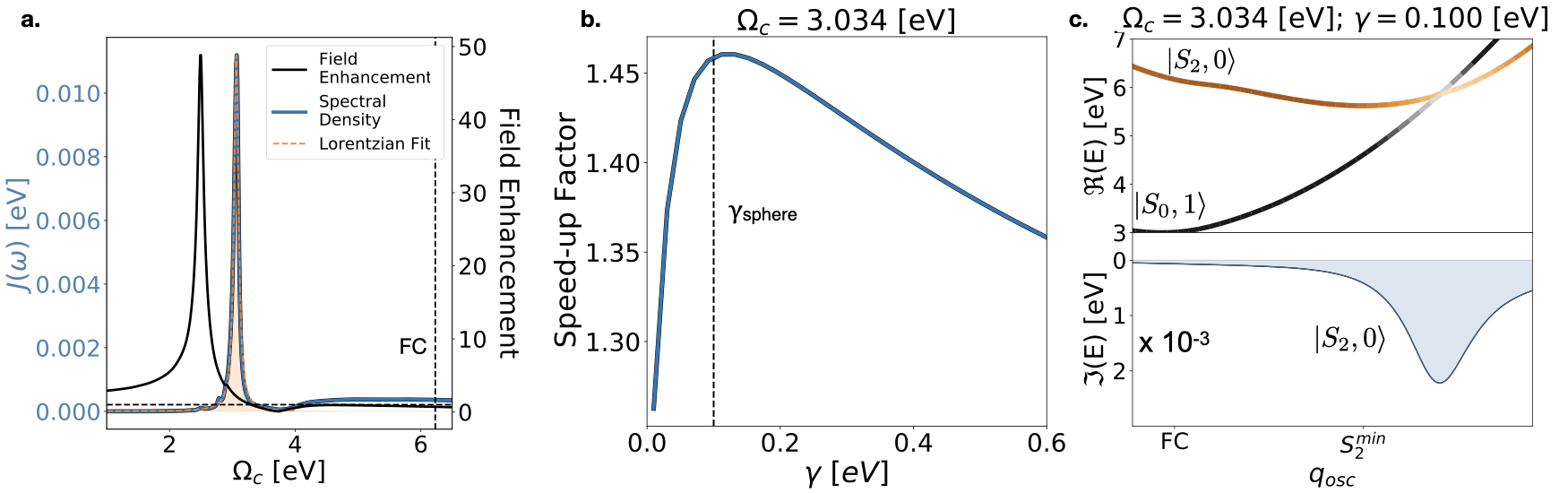}
  \caption{\textbf{Plasmonic pseudomode of a silver nanosphere of 30 nm diameter}|\textbf{a)} Lorentzian fit (dashed orange line) of the plasmonic pseudomode (full blue line) in a silver nanosphere, located centered at 3.0 eV. The agreement with the Lorentzian shows that the nanoparticle can be well-approximated by a single-mode cavity. The full black line in panel \textbf{a)} shows the field enhancement. The absence of field enhancement at FC (6.2 eV, vertical dashed line) guarantees that the presence of the nanoparticle does not enhance the absorption of the sample, which could potentially result in an augmented photodamage probability. \textbf{b)} The cavity decay rate associated to the silver nanosphere pseudo-mode ($\gamma$ 0.1 eV) also yields the maximum speed-up factor for the uracil photorelaxation when $\Omega_{cav}$ 3.0 eV. \textbf{c)} Real and imaginary parts of the polaritonic energy close to resonance. The splitting in the real part is zero, hence no transfer to the $\ket{S_0,1}$ state should be observed due to the coupling. Yet, a $\sim$1.5 speed-up of the reaction is still observed by effect of the imaginary contribution to the coupling on $\ket{S_2,0}$.}
  \label{fig:sphere}
\end{figure*}

The molecule-nanoparticle interaction can be described in terms of a pseudomode, which is an effective representation of an ensemble of independent harmonic oscillators\cite{garraway:pseudo}. For the present case, the pseudomode represents a manifold of plasmonic quasi-degenerate multi-pole modes embedded in a dielectric environment\cite{delga:pseudo}, which are coupled to the $S_0\rightarrow S_2$ transition.
In Figure \ref{fig:sphere}a we report the computed spectral density (blue continuous line) and the corresponding Lorentzian fit (orange dashed line). The good agreement between the spectral density and the fit shows that the nanoparticle is well approximated by a single pseudomode with resonance frequency $\Omega_c = 3.0$ eV and decay rate $\gamma = 0.1$ eV ($\sim$ 6-7 fs lifetime). We note that due to the proximity of the molecule to the sphere, this dominant mode corresponds to a combination of high-order multipole modes in the sphere \cite{delga:dipole}, and not the dipole resonance of the sphere at $\Omega \sim$ 2.5 eV (which is the only efficiently coupled to free-space radiation, as seen in the field enhancement in Figure \ref{fig:sphere}a).  We set the dipole moment to $\mu = 4.2$ D, that is the value of the uracil model at the relevant crossing point (see Figure \ref{fig:maps}c)\cite{regina1}. The light-matter coupling strength $g$ with such parameters measures $0.042$eV, which is consistent with the value of $e_{1ph}$=0.001 au considered in the numerical simulations throughout this paper.

In Figure \ref{fig:sphere}a we also show the enhancement of the field (full black line) intensity experienced by the molecule due to the nanoparticle, considering an incident plane wave. It is important to remark that at the FC frequency ($\Omega_{FC} \sim 6.2$eV) the field enhancement is slightly smaller than one, implying that the presence of the cavity does not improve the absorption efficiency. By reaching strong coupling with a localized photonic mode (Fig.\ref{fig:sphere}a), we are able to meet two apparently contradictory requirements to improve the uracil photoprotection mechanism: the introduction of an efficient photorelaxation channel and the absence of enhancement of the photoexcitation rate.

In Figure \ref{fig:sphere}b, we extract the section of the speed-up factor map (Figure \ref{fig:maps}b) at the pseudo-mode frequency, namely $\Omega_{cav} = $ 3.0 eV. Notice that there exists an optimal value of the decay rate which maximizes the speed-up factor. The optimal condition depends on the photon mode frequency, as it is given by a complex interplay between different factors such as the detuning from the optimal frequency, the speed of the wavepacket and the strength of the light-matter interaction. For example, the optimal decay rate $\gamma = $ 0.1 eV, (6-7 fs lifetime) observed for $\Omega_{cav} = $ 3.0 eV is larger than the one estimated to be the best for $\Omega_{cav} = $ 2.8 eV, \textit{i.e.} $\gamma = $ 0.05 eV ($\sim$ 12 fs). Notice that for the considered silver nanoparticle, the decay rate associated with the pseudo-mode $\gamma_{sphere}$ = 0.1 eV is very close to the maximum speed-up value, showing that optimal enhancement of the photorelaxation efficiency can be obtained with standard nanophotonic structures.

Finally, in Fig.\ref{fig:sphere}c we show the real and imaginary parts of the non-Hermitian polaritonic states obtained with the physical parameters of the silver nanosphere pseudomode. As the coupling strength is smaller than the
exceptional point value $g < \gamma/2$, there is no splitting in the real part of the polaritonic energies. Accordingly, no avoided crossing is observed and the polaritonic energies overlap with the bare PESs. In this case the systems is said to be in the weak-coupling regime, and it is not possible to observe a coherent excitation transfer between the states $\ket{S_2,0}$ and $\ket{S_0,1}$ within the cavity lifetime. However, the coupling with the pseudomode introduces a decay mechanism that is described in the non-Hermitian formalism by a localized complex potential, which is shown in the lower panel of Figure \ref{fig:sphere}c. Accordingly, the nanoparticle mediates an efficient channel that can speed up the photorelaxation process by $\sim 1.45$ times (as shown in Figure \ref{fig:sphere}b) even in the weak-coupling regime.

\subsection*{Conclusions}
In the present work, using 2D wavepacket dynamics calculations on lossy polaritonic PESs, we have shown how the coupling of uracil molecules with localized electromagnetic modes can be used to open an additional relaxation pathway, which is up to twice as efficient as the natural photoprotection mechanism hard-wired in the molecular structure. We have characterized the physical properties of the photonic device that optimizes the photorelaxation mechanism, and we have identified optimal conditions which do not require the implementation of complex nanophotonic structures. The highest efficiency is obtained at the limit between the weak and the strong coupling regimes. An important consequence emerges from these results: improving the nanocavity lifetime does not necessarily enhance the photoprotection efficiency.
Even more, the new relaxation pathway is already efficient when coupling simple metallic nanoparticles to the uracil molecule in the weak coupling regime. 
We show that a simple silver nanosphere embedded in a dielectric background can lead to a speed-up of about 50\% of the relaxation dynamics. Although a coherent transfer of population does not occur in the weak coupling regime, the photon mode introduces an effective complex potential which is sufficient to significantly improve the photorelaxation. 
Through the description the nanophotonic structure, we show that the mode under study does not enhance the absorption of the molecule in the photodamaging UVB excitation window for uracil. Consequently, the coupling introduces an additional photorelaxation channel without enhancing the photoexcitation efficiency, resulting in a purely photoprotective effect.
In conclusion, by merging chemically and physically accurate descriptions of molecules and nanoparticles, we have shown that photon decay can play an active role in the modification of chemical reaction rates induced by nanophotonics structures, in the context of molecular polaritonics. Our study paves the way to the use of lossy photonic devices as a tool to tailor photorelaxation channels and to selectively inhibit reaction pathways. Due to its simplicity, the setup proposed for the silver nanosphere and uracil can be feasibly implemented experimentally with current nanophotonic technology.

\subsection*{Methods}

\paragraph{Molecular calculation} The isolated uracil potential energy surfaces are computed at MRCI(12,9))/cc-pVDZ level\cite{regina1} with an active space of 12 electrons in 9 orbitals, with single excitations allowed out of the active space into the virtual space.
The propagation was performed by relying on a finite elements discrete variable representation (FEDVR\cite{fedvr}) spatial grid, including 11 spline basis functions for each grid point to represent the nuclear wavefunction. In the propagation, the non-adiabatic coupling vectors are evaluated at SA-CASSCF level within the same active space on the same FEDVR grid\cite{regina2}. As the WP reaches $S_1$ from $S_2$ through the conical intersection, an absorbing potential gradually set along the slopes of the $S_1$ state acts to decrease the WP norm. By these means, we mimic the fast internal conversion from the $S_1$ state, together with avoiding unphysical behaviours due to the WP travelling back to $S_2$.

\paragraph{Light-matter Hamiltonian}

We consider a model composed of a uracil nucleobase individually coupled to a single quantum optical mode, of frequency $\Omega_c$ and decay rate $\gamma$. Let us first describe the system full Hamiltonian, we will discuss later how the dissipation is included in our model. The total  Hamiltonian is given by three main components $\hat H = \hat H_{mol} + \hat H_{\rm cav} + \hat H_{int}$. The bare molecular energy can be written as:
\begin{equation}
\hat H_{mol} = \hat T_{nuc  } + \sum_i V_i(\mathbf{R}) \ket{S_i}\bra{S_i} + \sum_{i,j} \bm{G_{ij}}(\mathbf{R}) \ket{S_i}\bra{S_j},
\end{equation}
where $\hat T_{nuc}$ is the nuclear kinetic energy operator, $i$ and $j$ denote the bare molecule electronic states in the adiabatic representation. $V_i(\mathbf{R})$ are the potential energy surfces (PESs), and the term $\bm{G_{ij}}(\mathbf{R})$ gathers the non-adiabatic couplings vectors which correct the Born-Oppenheimer approximation. 
The photonic contribution to the Hamiltonian is written as the quantized electromagnetic field Hamiltonian:
\begin{equation}
\hat H_{\rm cav} = \Omega_c \hat a^\dagger\hat a, 
\end{equation}
in terms of creation $\hat a^\dagger$ and annihilation $\hat a$ operators, where $\Omega_c$ is the cavity frequency. Finally, the light-matter interaction term is given by
\begin{equation}
 \hat H_{int} = \sum_{i,j} g_{ij}(\mathbf{R}) \ket{S_i}\bra{S_j}\left(\hat a^\dagger + \hat a \right),
\end{equation}
 where  the interaction strength $g_{ij}(\mathbf{R}) =  \bm{\mu_{ij}}(\mathbf{R})\cdot \mathbf{E_{1ph}}$ is given by the scalar product of the dipole transition moments $\bm{\mu_{ij}}(\mathbf{R})$ with the electric field generated by a single photon and   polarized along $\bm{\lambda}$, that is $\mathbf{E_{1ph}}=\bm{\lambda} e_{1ph}$.

The full space to propagate for the cavity-molecule system is composed by the manifold of the $\ket{S_i,p}$ states, where $i$ is the electronic state index and $p$ is the cavity occupation number. For the molecule, we restrict to the three electronic states directly involved in the free evolution dynamics, namely $\ket{S_2,0}$, $\ket{S_1,0}$ and $\ket{S_0,0}$. We shall now discuss some reasonable approximations for the cavity-molecule coupling to reduce the complexity of the model and to cut the computational cost.
 
First, we perform the rotating-wave approximation and hence neglect the far off-resonant terms, namely the coupling to the states $\ket{S_i,p\geq1}$. Among the off-resonant states, the lowest in energy is $\ket{S_1,1}$, which sits at least 2 eV above the $\ket{S_2,0}$. This difference is well beyond the relevant window for the dynamics occurring on $\ket{S_2,0}$.
 Second, we assume that the cavity mode is occupied by at most one photon, which is always the case under rotating-wave approximation if the cavity mode itself is not externally driven.  Accordingly, in the reduced model the potential energy landscape is composed of four relevant PESs, corresponding to the  states  $\ket{S_0,0}$, $\ket{S_1,0}$,  $\ket{S_2,0}$, and $\ket{S_0,1}$. In absence of any light-matter interaction, the states $\ket{S_i,0}$ correspond to the potentials $V_i(\mathbf{R})$, while the state $\ket{S_0,1}$ has the energy $V_0(\mathbf{R}) + \Omega_c$, that is  the $S_0$ PES lifted by the cavity frequency (Figure \ref{fig:sc_dynam}a and \ref{fig:sc_dynam}b). The light-matter interaction for the present case is given by the Hamiltonian:
\begin{equation}
\label{lm_coupling}
\hat H_{int}  = g_{02}(\mathbf{R})\Big(\ket{S_0,1}\bra{S_2,0} +  \ket{S_2,0}\bra{S_0,1} \Big),
\end{equation}
and so it can induce transitions between the states $\ket{S_2,0}$ and $\ket{S_0,1}$.  Let us now discuss how dissipative effects are included in our formalism.

\paragraph{Non-Hermitian propagation}
Cavity losses can be formally taken into account using the Lindblad master equation, which is based on the assumption that the cavity mode is weakly coupled to a Markovian bath. At zero temperature, the evolution of the system density matrix is given by,
\begin{equation}
\label{ME}
\dot \rho = -i \left[\hat H, \rho \right] + \gamma \hat a \rho \hat a^\dagger - \frac{\gamma}{2} \left(\hat a^\dagger \hat a \rho + \rho \hat a^\dagger \hat a \right),
\end{equation}
where $\gamma$ is the photon decay rate. Notice that the term $\hat a \rho \hat a^\dagger$ induces incoherent transitions $\ket{S_i, p} \rightarrow \ket{S_i, p-1}$, which in our reduced subspace corresponds only to the transition $\ket{S_0, 1} \rightarrow \ket{S_0, 0}$. Given that we want to focus on the population that is leaking out from the $S_2$ electronic state, we can reduce further the computational space, keeping only into account the states $\ket{S_1, 0}$, $\ket{S_2, 0}$ and $\ket{S_0, 1}$.
In this subspace the term $\hat a \rho \hat a^\dagger$ vanishes and can be dropped. We can then rewrite the master equation \eqref{ME} in terms of a non-Hermitian Hamiltonian~\cite{Visser95}, $\dot \rho = -i \left(\hat H_{NH}\rho - \rho \hat H_{NH}^\dagger \right)$ where $\hat H_{NH} = \hat H - i\frac{\gamma}{2}  \hat a^\dagger \hat a$.  In this non-Hermitian formalism, photon losses are then  kept into account by the loss of norm of the system state during the time-evolution. 
The system evolution can be then calculated solving the non-Hermitian Schroedinger equation \emph{for state vectors}, instead of using the full Lindblad master equation  \emph{for density matrices}.

\singlespacing
\bibliography{biblio.bib}
\onehalfspacing

\paragraph*{Data Availability}
The codes and data devised and used in the realization of the present work are available from the corresponding author upon reasonable request.

\paragraph*{Conflict of Interest}
The authors declare no competing financial interest.

\paragraph{Acknowledgements}
The authors thank Stefano Corni (University of Padova) for fruitful scientific discussion.
This work has been funded by the European Research Council through grants ERC-2016-StG-714870 (S. Felicetti, J. Feist, J. Fregoni) and ERC-2015-CoG-681285 (J. Fregoni, PI Stefano Corni), and by the Spanish Ministry for Science, Innovation, and Universities – Agencia Estatal de Investigación through grants RTI2018- 099737-B-I00, PCI2018-093145 (through the QuantERA program of the European Commission), and MDM-2014-0377 (through the María de Maeztu program for Units of Excellence in R\&D). TS and RdVR gratefully acknowledge the DFG Normalverfahren.
SR gratefully acknowledges financial support by the International Max Planck Research School of Advanced Photon Science (IMPRS-APS).

\paragraph*{Authors Contribution}
SF, JFr and JFe initiated the project. TS performed the calculations on the molecule. TS, SR and RdVR provided the uracil data and actively contributed to devising the analysis and debugging. SF and JFr implemented the codes, performed the strong coupling and nanoparticle calculations and contributed equally to the realization of the present work. All authors contributed to the final version of the manuscript.
\end{document}